\documentclass[%
 reprint,
 amsmath,amssymb,
 aps,
prd,
]{revtex4-2}

\usepackage{graphicx}
\usepackage{dcolumn}
\usepackage{bm}

\usepackage{hyperref}
\usepackage{cleveref}
\usepackage{booktabs}
\usepackage{multirow}
\usepackage{tabularx}
\usepackage{siunitx}
\DeclareSIUnit{\solarMass}{M_{\odot}}\DeclareSIUnit{\solarMass}{M_{\odot}}

\begin{document}


\title{Robust statistical significance for Gravitational Wave inference}

\author{Alexandre S. Göttel}

\email{alexandre.goettel@nottingham.ac.uk}
\affiliation{
School of Mathematical Sciences, University of Nottingham, Nottingham NG7 2RD, United Kingdom
}

\date{\today}

\begin{abstract}
The field of gravitational-wave data analysis heavily relies on Bayes factors to make astrophysical statements about detected signals. This is true on an event level, for example when discussing the potential detection of orbital precession or eccentricity, as well as on a population level. For the latter, a main limiting factor is the growing bias resulting from accumulating per-event uncertainties related to stochastic sampling. In both cases, the computational cost of calculating Bayes factors severely limits the astrophysical usefulness of the results: Though Bayes factors robustly quantify the data's preference towards one model, one cannot generally determine how likely an equally large value is to arise under the null hypothesis.
This paper applies evidence networks, amortized neural estimators which can infer Bayes factors after learning the distinguishability between two datasets, to the aforementioned problems.
This makes it possible to construct empirical Bayes factor distributions under different signal assumptions, providing a calibration between observation and discovery significance.
Using simulated Gaussian noise, we apply this method to a search for orbital precession in gravitational-wave data and achieve compatible results to current state-of-the-art approaches that require stochastic sampling. We further show that high precision of the achieved results leads to recovering injected isotropic-spin population fractions without discernible bias for populations of up to $10^4$ events on a simplified example.
More generally, because the method does not require explicit likelihood evaluations, this lays a clear path towards a likelihood-free framework that does not rely on the assumption of Gaussian and stationary noise.

\end{abstract}

\maketitle

\section{Introduction} \label{sec:intro}
Gravitational-wave (GW) data analysis has from before the first detection to today been dominated by Bayesian approaches. More specifically, the parameter estimation of all LIGO-Virgo-KAGRA (LVK) catalogues~\cite{ligo2015advanced,VIRGO:2014yos,akutsu2021overview,gwtc5} were performed using a \textsc{Bilby}-implementation of the \textsc{Dynesty} Nested-Sampling algorithm~\cite{bilby_paper,dynesty}. The primary design goal of nested sampling is to estimate the so-called \textit{evidence} $\mathcal{Z}$, obtained by integrating the prior-weighted likelihood over the entire, typically high-dimensional, parameter space. It encodes all necessary information for model comparison, and is mostly used to calculate Bayes factors which, assuming equal prior odds, equals the posterior odds at which the data favours one hypothesis over another.

\noindent The calculation quickly suffers from sampling uncertainty, and in the case of all compact binary coalescence (CBC) parameter estimation by the LVK relies on the assumption that the detector noise is perfectly Gaussian and stationary. Both of these assumptions are known to be false, as for example at least 20\% of GW detections contain transient noise artifacts~\cite{GWTC4}, but deliver `good-enough` results for most events. 

\noindent The high costs of calculating evidences with nested sampling, typically involving $\mathcal{O}(10^{1-2})$ hours of wall-time and $\mathcal{O}(10^{7-8})$ likelihood evaluations for a single evidence value, leads to another important limitation: it is not generally possible to evaluate the probability of obtaining an equally extreme Bayes factor under a null hypothesis. This severely limits our ability to make robust statistical statements.

\noindent A related scaling problem occurs for population inference. The prohibitive costs of a joint event-and-population parameter inference resulted in the widespread adoption of hierarchical Bayesian Inference for gravitation-wave population inference~\cite{thrane2020introduction}. Because these methods rely on the Monte-Carlo integration of samples from a large number of event-level stochastic inference results, accumulating finite-sample and Monte-Carlo errors result in biases which are already a huge obstacle to the field~\cite{talbot2023growing,heinzel2025not} and will quickly grow more problematic with gravitational-wave catalogue sizes.

\noindent This paper uses a new method~\cite{Evidence} that uses machine learning to directly estimate Bayes factors from data by exploiting underlying statistical distinguishability.
Amortized, the method delivers estimates in instants as opposed to hours or weeks with nested sampling.
We apply this approach to a search for orbital precession in gravitational-wave data. Precession is a direct marker of astrophysical merger progenitors, and has been notoriously difficult to measure even when assuming Gaussian noise. We first introduce carefully developed procedures designed to obtain reproducible results, and validate the resulting networks using coverage tests, consistency between independent analyses, and direct comparisons with nested sampling. In idealised Gaussian noise, where independent validation against nested sampling is possible, we show that the trained networks allow the Bayes factor distribution under a null hypothesis to be sampled directly, delivering compatible results to costly state-of-the-art approaches.

\noindent Finding that our results are also substantially more precise than those obtained with stochastic samplers, we further apply our networks to a population-level analysis. We show that we can accurately recover the isotropic fraction with no discernable bias for up to at least $10^4$ events. By bypassing the repeated use of finite event-level samples, this calculation remains inexpensive as the catalogue sizes increase. Most importantly, the method does not require a formal definition of the likelihood or priors. Any working implementation can thus in principle be extended to learn directly from any available data, including real (non-Gaussian, non-stationary) interferometer noise. Contrary to nested sampling, this would allow robust statistical inference on real detector noise.

\Cref{sec:theory} establishes the method in the context of GW precession measurements. \Cref{sec:methods} describes the training and validation procedures with modifications specific to GW data analysis.
\Cref{sec:discussion} presents the calibration of Bayes factors with statistical significance and the application to population inference.

\section{Background} \label{sec:theory}
\subsection{Evidence networks} \label{sec:theory:evidence}
In essence, evidence networks are simple binary classifiers that are trained to differentiate between two datasets. The idea is that with a clever choice in the loss function, one can calibrate the optimal network output, given observed data $x_\mathcal{O}$, to be a trivial transform of the Bayes factor $\mathcal{B}$ between the two models. Because the output of such a network is just a single number, one essentially bypasses the high-dimensional calculations (and accompanying uncertainties and numerical instabilities) that are typically required, for example when using nested sampling. This is possible because the loss exploits the statistical properties of the Bayes factor as a discriminator in data-space directly, instead of trying to derive it through integration. The drawback is that such losses necessarily include exponential terms, which because of the behaviour of their gradients are notoriously difficult to handle with deep learning.

It is worth noting that in theory, the widely-used binary cross-entropy loss could also be used. After all, it outputs probabilities that can be converted to Bayes factors. However, in practice, since neural networks can typically only handle numbers of $\mathcal{O}(1)$ when back-propagating gradients over several layers, one can't expect them to perform reliably in probability spaces where one model is strongly preferred over another. Additionally, standard classifier training generally struggles with the precision needed for meaningful model comparison, and often produces overconfident results~\cite{Calibration}. While this is not necessarily a problem for pure classification tasks, it renders it practically impossible for such approaches to be accurate beyond an overly narrow range of Bayes factors.

The work in this paper is based on~\cite{Evidence}, where methods were explored to counteract this problem. Specifically, we use the `leaky` parity-odd power (l-POP) exponential loss function. It is defined as:
\begin{equation} \label{eq:loss}
    \mathcal{L}(f(x), m) = e^{(\frac{1}{2} - m)\mathcal{J}_\alpha(f(x))},
\end{equation}
where $m \in [0,1]$ denotes the source of the data $x$, and $f(x)$ represents the output of the network applied to the data. The l-POP transform $\mathcal{J}_\alpha$ is defined as:
\begin{equation} \label{eq:j_alpha}
    \mathcal{J}_\alpha(x) := x + x |x|^{\alpha - 1}.
\end{equation}
Though one could replace $\mathcal{J}_\alpha$ with any monotonic function and still obtain valid results, we find, in line with the original publication, that using it leads to more stable network training. The reason for using this exponential loss is, as outlined in~\cite{Evidence}, that an \textit{optimal} network $f^*(x)$, trained to reach a global minimum on this loss, will output a simple transform of the Bayes factor $\mathcal{B}$ for observed data $x_\mathcal{O}$, with:
\begin{equation}
    \log \mathcal{B} = \mathcal{J}_\alpha(f^*(x_\mathcal{O})).
\end{equation}

\subsection{General-relativistic Precession in GW data}
In binary black hole (BBH) systems, precession stems from the misalignment between the orbital angular momentum and the total spin of the individual black holes. The effects of general-relativistic precession have long been studied, and its possible imprints on gravitational waves are generally well understood, at least in theory~\cite{precession,green2021identifying}.

\noindent Precession is of particular astrophysical interest, as its imprints on gravitational waves are a clear indication of the formation channel of the binary. Indeed, while many models of BBH formation exist, it is commonly accepted that the two main expected categories are binaries formed with a common envelope, and those formed in the presence of dynamical interactions in high-density environments~\cite{gwtc5pop}. The former implies friction that would force the component spins to align with the orbital angular momentum, while in the latter the spins are expected to be isotropic. A confident detection of precession, along with the properties of said system, would therefore inform us on the fraction of events from each formation channel, and greatly impact models on the role of clusters in the
BBH population~\cite{antonini2025inferring}. It would also help to understand pair-instability supernov\ae\, which would in turn impact models of stellar evolution, notably given the sensitivity of the pair-instability mass gap to the $^{12}\text{C}(\alpha,\gamma)^{16}\text{O}$ reaction rate~\cite{woosley2017pulsational}.

There is to date but a single GW event for which substantial evidence for precession has been claimed: GW200129~\cite{GWTC3,precessionRare,Hannam2022}, and even in this case the results are disputed.
In particular, while~\cite{Hannam2022} find 1:25 000 odds in favour of precession,~\cite{payne2022curious} finds that the difference in the data between precessing and non-precessing hypotheses is lower than the statistical uncertainty resulting from methods used to remove known excess noise from the data. However, other studies using different methods of noise mitigation~\cite{macas2024revisiting,hoy2026parametric} find compatible results with~\cite{Hannam2022} (though neither quantifies the significance of a potential measurement of precession). In any case, it is noteworthy that the only event with such strong claims is also known to have excess noise in precisely the frequency range in which the evidence for precession originates~\cite{payne2022curious}.

\noindent Generally speaking, the main obstacle behind the detection of precession is that its leading order effect on GW waveforms as measured from Earth vanishes when the BBH is face-on. Such BBH systems have a higher signal-to-noise ratio (SNR) and are therefore the most common type of detected systems. When factoring in knowledge from population studies, ~\cite{precessionRare} finds that only about 1 in 50 events can be expected to have a \textit{significant} imprint of precession. While the meaning of \textit{significant} is somewhat up to interpretation since the data includes GW200129, even in optimistic scenarios the evidence for the presence of measurable precession is quoted to be less than $3\sigma$ when considering all events in GWTC-3~\cite{GWTC3}.

\section{Methods} \label{sec:methods}
\subsection{Data} \label{sec:methods:data}
As mentioned in \Cref{sec:theory:evidence}, evidence networks are classifiers trained to differentiate between two datasets, calibrated such as to output (up to a trivial transform) the Bayes factor between the models used to generate said datasets.
We train two such networks. The first, \textit{Scenario I}, compares an isotropic-spin model (A) with an aligned-spin model (B). Both use the \textsc{IMRPhenomXPHM}\cite{xphm} waveform approximant and the default LVK isotropic-spin priors\cite{GWTC4}, except that model B fixes the component spins to align with the orbital angular momentum. We restrict the mass and distance priors to be compatible with published GW231123 posteriors~\cite{abac2025gw231123} and the \textsc{NRSur7dq4}~\cite{nrsur} approximant, allowing these analyses to be reused in future work. More details on the priors can be seen in \Cref{tab:priors}.

\noindent We also train a network for \textit{Scenario II}, in which both models use identical priors, but data are generated using either (A) \textsc{IMRPhenomXPHM}~\cite{xphm} or (B) \textsc{IMRPhenomPv2}~\cite{pv2_I,pv2_II} (which does not model multipoles beyond the 22 mode). This scenario was designed to generate data with low Bayes factors in order to simplify network training, and such that \textsc{Bilby} analyses using the scenario will finish in minutes. This will be useful in later sections of this paper where $\mathcal{O}(10^2)$ Bayes factors are computed with \textsc{Bilby} to validate the networks. Given the computational costs associated with nested sampling, this would not be feasible for \textit{Scenario I}.

\begin{table}[t]
    \centering
    \begin{tabular}{lccc}
        \toprule
        & \multicolumn{2}{c}{\textbf{Scenario I}} 
        & \textbf{Scenario II} \\
        \cmidrule(lr){2-3}
        \textbf{Parameter}
        & \textbf{Shared}
        & \textbf{Model dependent}
        & \textbf{Models A/B} \\
        \midrule

        \multicolumn{4}{l}{\textit{Intrinsic parameters}} \\[2pt]

        $\mathcal{M}$ 
        & $85$--$195$ 
        & --- 
        & $80$--$120$ \\

        $q$ 
        & $1/6$--$1$ 
        & --- 
        & $1/8$--$1$ \\

        $M$ 
        & $60$--$400$ 
        & --- 
        & --- \\

        $a_1$ 
        & default 
        & --- 
        & $0$ \\

        $a_2$ 
        & default 
        & --- 
        & $0$ \\

        $\theta_1$ 
        & --- 
        & A: default;\quad B: $0$
        & $0$ \\

        $\theta_2$ 
        & --- 
        & A: default;\quad B: $0$
        & $0$ \\

        $\phi_{JL}$ 
        & --- 
        & A: default;\quad B: $0$
        & $0$ \\

        $\phi_{12}$ 
        & --- 
        & A: default;\quad B: $0$
        & $0$ \\

        \addlinespace
        \multicolumn{4}{l}{\textit{Extrinsic parameters}} \\[2pt]

        $D_L$
        & $10^3$--$10^4$
        & ---
        & $4000$ \\

        $t_c$
        & default
        & ---
        & default \\

        $\phi_c$
        & default
        & ---
        & default \\

        $\delta$
        & default
        & ---
        & $0.3$ \\

        $\alpha$
        & default
        & ---
        & $4.4$ \\

        $\theta_{JN}$
        & default
        & ---
        & $0.75$ \\

        $\psi$
        & default
        & ---
        & $1.7$ \\

        \bottomrule
    \end{tabular}
    \caption{Parameters used to generate the simulated gravitational-wave data. ``Default'' denotes the standard uninformative priors used in LIGO--Virgo--KAGRA catalogue analyses~\cite{gwtc5}. Parameters listed under ``Shared'' are identical for both competing models. Fixed parameters are given explicitly. In \textit{Scenario II}, both models use the same prior.}
    \label{tab:priors}
\end{table}


\noindent Data from all models are generated in the frequency domain from \SIrange{20}{224}{\hertz}, a resolution of \SI{0.25}{\hertz}, and a reference frequency of \SI{10}{\hertz}. Coloured Gaussian noise is added using a PSD generated with the ``aLIGO'' configuration of \textsc{pygwinc}
~\cite{pygwinc} and in the Hanford detector~\cite{ligo2015advanced} only.

\subsection{Training} \label{sec:methods:train}
Neural network training famously works best when input and output values are of $\mathcal{O}(1)$. In the field of gravitational-wave data analysis, it is not rare to see $\log\mathcal{B} \sim \mathcal{O}(10^2)$, for example when calculating the odds of there being a gravitational wave in noisy data. However, one can quickly see that even for $\log\mathcal{B} = \pm20$, loss values (see \Cref{eq:loss}) will reach $\mathcal{O}(10^{\pm 8})$. This is unavoidable since exponential terms are necessarily part of a loss that is used to output log Bayes factors, and represents a huge challenge for backward propagation even over a few layers. 
This section discusses the methods used to make training possible and reliable in this difficult loss landscape.

\noindent The first and most obvious step to avoid gradient explosion is to make sure that the training data does not contain examples with a corresponding $\log\mathcal{B}$ outside of a roughly $\pm 50$ range where float precision will start failing in the exponential terms. For problems leading to such extreme values of $\log\mathcal{B}$, it is anyway questionable whether it is sensible to compare the two models in the first place.

\noindent A series of steps can then used to directly affect gradient updates: 1. gradient clipping, meaning that after backpropagation all gradients above a certain threshold are capped at the threshold value before the optimisation step occurs 2. batch normalisation, where layer inputs are normalised in each batch during training and an average normalisation is applied during validation 3. initialising the biases of the final layer to zero, and the weights in a $\pm10^{-3}$ range. This dampens gradients in the initial training steps where the learning rate is usually highest, preventing erratic trajectories following a bad initialisation from leading to an unrecoverable state (which can easily occur in areas where the loss saturates because of the exponential term). Finally, 4. using an optimiser with a global approach to weight updates, like stochastic gradient descent, instead of one using adaptive methods.

\noindent These measures stabilised individual training runs, but independently trained networks still produced inconsistent Bayes-factor estimates and largely uninformative outputs outside a narrow range around $\log\mathcal{B} = 0$.
This is easily understood in the context of both scenarios studied in this paper. Since the majority of data points have low evidence, the optimiser will struggle to when moving into the tails of the distributions. This problem was solved by using a so-called GrowNet~\cite{grownet} approach. This consists of training a secondary network, identical to the original one with the exception that, next to the training data, it also has access to the output of the penultimate layer of the pre-trained network.
In theory, if the first network finds a solution that will satisfy the bulk of the data, then the second iteration can focus on applying corrections to it, thereby reaching the next leading order. This was found to greatly stabilise the output of the networks and reliably make them reach lower loss values (see \Cref{sec:methods:val}).\\

\noindent It is worth mentioning that besides the exponential gradients, another reason for difficulties during training is that, in order for the networks to actually output Bayes factors instead of only acting as classifiers, the training data must be generated by exactly following the prior. One can thus not generally apply transforms that affect data space to make training easier unless they can be shown to be orthogonal to the distinguishability of the datasets. Similarly, embeddings must be sufficiently flexible to preserve the information required for model comparison. Otherwise, the network estimates Bayes factors for the compressed representation, which may differ from those for the original data. Separately, the training distribution must provide sufficient coverage in rare regions of the parameter space needed to constrain the tails of the Bayes factor distribution.

All network training in this paper is performed using the noisy data described in ~\Cref{sec:methods:data} by concatenating $4\cdot 10^6$ complex amplitudes and angles, and using a relatively simple architecture (see \Cref{tab:architecture}) owing to the simple binary-classifier nature of the networks.
The training used a learning rate of \(6 \cdot 10^{-4}\), momentum of 0.94, weight decay of \(4 \cdot 10^{-4}\), a gradient clip of 0.16, and a batch size of 500.

\begin{table}[h]
    \centering
    \begin{tabular}{l|c}
        \toprule
        \textbf{Layer} & \textbf{Contents} \\
        \hline
        Input Layer & 512 neurons \\
        (x4) ResNet Block & 3 layers, 128 neurons each \\
        Output Layer & 1 neuron \\
        \hline
    \end{tabular}
    \caption{Architecture of the evidence networks used in this paper. All activation functions are \textsc{leaky rectified linear unit}~\cite{pytorch} except the last one which uses \Cref{eq:j_alpha}.}
    \label{tab:architecture}
\end{table}

\subsection{Validation} \label{sec:methods:val}
As mentioned in \Cref{sec:theory:evidence}, an optimally trained evidence network will output a trivial transform of Bayes factors. However, there is never a guarantee that a network has indeed reached a global minimum, even when its validation loss has reached a plateau. It is therefore important to use other means of validation. This section outlines \textit{blind coverage testing}, adapted from~\cite{Evidence}, and discusses further training and network consistency checks.

\begin{figure}
    \centering
    \includegraphics[width=\linewidth]{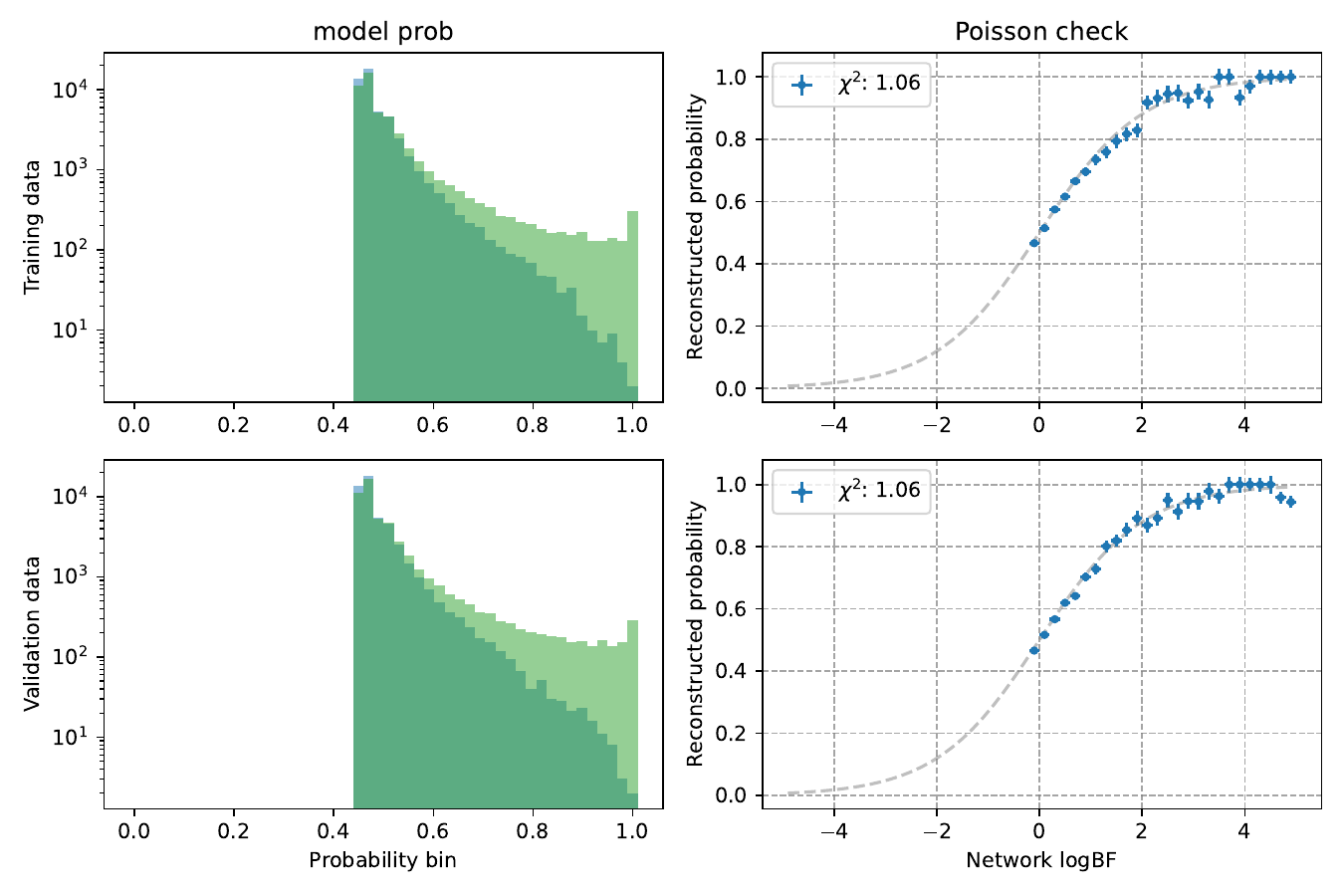}
    \caption{Blind coverage and $\chi^2$ tests using $10^5$ waveforms. \textit{upper}: using training data \textit{lower}: using validation data \textit{left}: histogram of $p(M_A|x)$ values (see \Cref{eq:prob}) where green and blue denote data from models A and B, respectively. Because model B is a subset of model A, one would not expect strong evidence for model B, hence the lack of values below about 0.5 \textit{right}: the data points are the ratio between the numbers in each histogram bin with Poisson-derived uncertainties, and the dashed curve shows \Cref{eq:prob}. As one can see, the output conforms to expectation.}
    \label{fig:covtest}
\end{figure}

\noindent Blind coverage testing makes use of the statistical properties of the Bayes factor, namely that for equal prior odds it is equal to the odds ratio. It follows that:
\begin{equation} \label{eq:prob}
    p(M_A|x) = \frac{\mathcal{B}}{1 + \mathcal{B}},
\end{equation}
where $p(M_A|x)$ is the probability of model A being true given the data $x$. Using validation data (i.e. data that a trained network has not \textit{seen}), one can thus simply use this equation to calculate $p(M_A|x)$ for every data point to construct a histogram. When using the same number of data points from each model, one would expect that in each bin, the relative number of entries from each model corresponds to the predicted probability. By assuming the numbers each bin to be independent and Poisson distributed, one can perform a $\chi^2$ test where the number of degrees of freedom is the number of bins. The result of this test for Scenario I can be seen in \Cref{fig:covtest}.

\begin{figure}
    \centering
    \includegraphics[width=\linewidth]{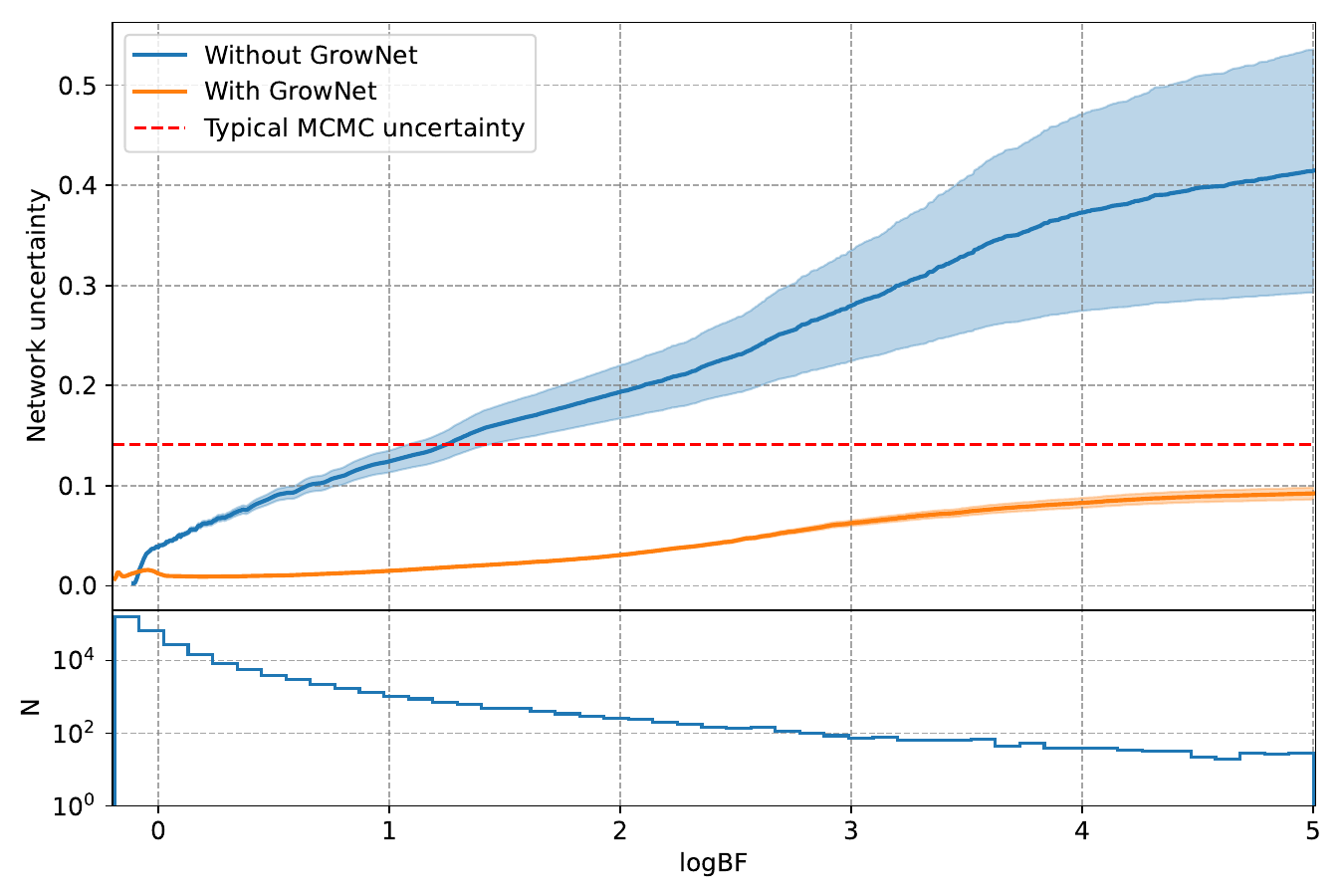}
    \caption{With and without GrowNet (see \Cref{sec:methods:train}), the standard deviation between five trained evidence networks as a function of the Bayes factor. The band shows the uncertainty on the estimated sample standard deviation, and the red line at $\sqrt{2}/10$ shows a typical uncertainty on log Bayes factors obtained from nested sampling.}
    \label{fig:consistency}
\end{figure}

\noindent While \textit{necessary}, this test is not \textit{sufficient} to prove that the networks have correctly converged.
A further validation test thus compares the output of separate networks trained on the same input data. Should the training procedure be valid, one would expect the networks to converge to the same result. For this test, N networks are separately trained with the same pre-determined procedure and on the same training data. The standard deviation of the outputs is then calculated for every waveform in the validation data and neighbouring values (in $\log\mathcal{B}$) are averaged for visibility. \Cref{fig:consistency} shows the output of this test, again for Scenario I. One can see that the GrowNet approach (see \Cref{sec:methods:train}) has a significant impact on training consistency. In this example, while uncertainties predictably increase in areas where there is less training data, it is nevertheless clear that the networks are consistent over the entire considered range of Bayes factors, and that their precision outperforms nested-sampling approaches by up to an order of magnitude. It is worth mentioning that this test can also be used to estimate the uncertainty of individual networks.\\

\begin{figure}
    \centering
    \includegraphics[width=\linewidth,trim={0.7cm 0 1.5cm 0},clip]{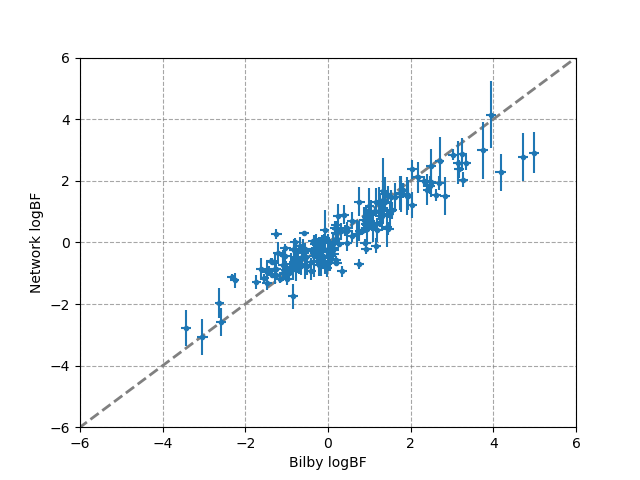}
    \caption{200 log Bayes factors obtained with the evidence networks against log Bayes factors obtained with \textsc{Bilby}. The y errors are estimated by training 5 networks and calculating the standard deviation between their outputs. As one can see, the data mostly fits the dashed $x=y$ line.}
    \label{fig:x-equal-y}
\end{figure}

\begin{figure}
    \centering
    \includegraphics[width=\linewidth,trim={0.9cm 1cm 2cm 1cm},clip]{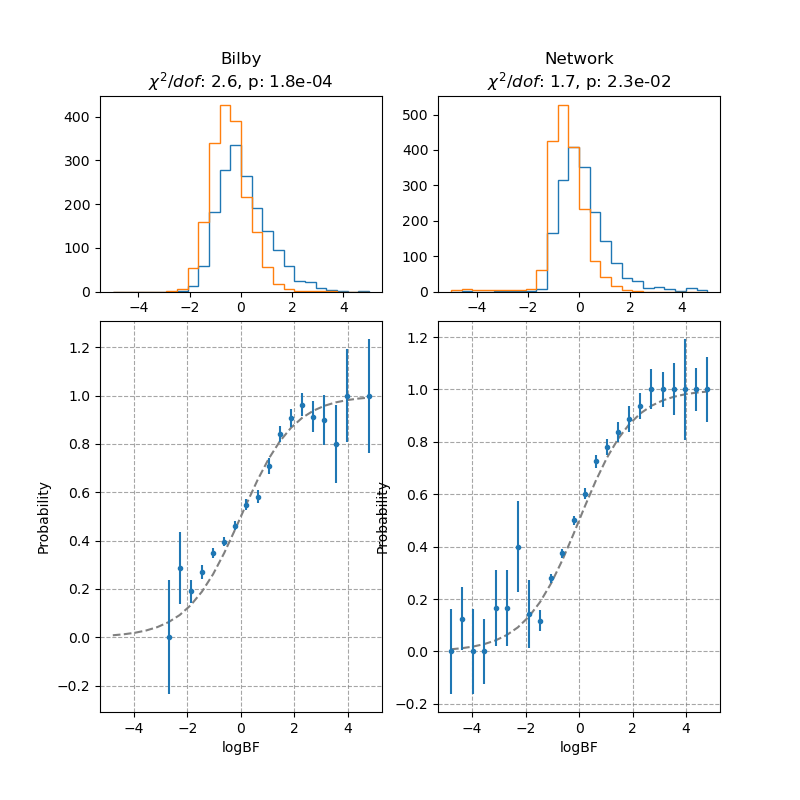}
    \caption{Blind coverage test using 200 waveforms \textit{left}: \textsc{Bilby} and \textit{right}: evidence networks. The histograms on top show the logBF distribution from the two models of scenario II, and the curves show the $\chi^2$ test (see \Cref{sec:methods:val}). One can clearly see that although both approaches give similar overall results, that the networks outperform \textsc{Bilby} on this problem.}
    \label{fig:covtest-bilby}
\end{figure}

We further compare the output from the networks to established methods. Being prohibitively expensive using problems like \textit{Scenario I}, we perform this comparison with \textit{Scenario II}. This is a simple four-parameter inference in which the free parameters are chirp mass, mass ratio, geocentric time, and phase. Bayes factors from \textsc{Bilby} were obtained using waveforms from the validation data. Since one inference takes about \SI{5}{\min} using $\textsc{Bilby}$ (\SI{10}{\min} per Bayes factor), it would still not be feasible to estimate $10^5$ Bayes factors as was done in \Cref{fig:covtest}. Instead, 200 Bayes factors were calculated with \textsc{Bilby} and compared to the average output of five evidence networks, see \Cref{fig:x-equal-y}, with uncertainties estimated from their sample standard deviation as mentioned above. As one can see, the results mostly agree, with a few exceptions where the deviation is significant given the low uncertainty between the network results, though it never deviates by more than 2. Repeated training with different settings showed that these differences are persistent \textit{i.e.} that the output from the networks is very consistent. Another possibility is that the uncertainty from \textsc{Bilby} (when run with a default \textrm{dlogz=0.1}) is underestimated. It is important to note that nested sampling also cannot generally obtain valid evidence values, as it can always miss sharply peaked modes in the likelihood structure during the integration process.
\Cref{fig:covtest-bilby} puts this hypothesis to the test, and one can indeed see that under a blind coverage test (see above), the networks outperform \textsc{Bilby}. This is consistent with~\cite{Evidence} where it was found that evidence networks outperform likelihood-based approaches, at least on a toy problem.

\section{Discussion} \label{sec:discussion}
\subsection{Precessing SNR} \label{sec:discussion:rhop}
The \textit{precessing SNR} ($\rho_p$) was proposed in~\cite{rhop,Fairhurst:2019srr} to help make statements about the detectability of precession. The idea is, having found that over the majority of the BBH parameter space the first two terms of the leading order precessing harmonics are the most significant, and given that both must be measurable for precession to be detected, one can simply define the minimum SNR between those harmonics as that with which precession can be detected. The main benefit of this approach is that the null hypothesis leads to a clearly defined background distribution. Indeed, from noise alone one would expect $\rho_p^2$ to follow a $\chi^2$ distribution with $k=2$ degrees of freedom. Thus, by using the samples from a nested sampling run in a precessing parameter space, one can obtain a significance against the hypothesis that the observed precessing SNR is due to noise. By additionally using samples from a nested sampling run restricted to the aligned-spin hypothesis, one can further inform the prior on $\rho_p$ to obtain a more accurate significance~\cite{green2021identifying}.

\noindent As argued in~\cite{rhop,green2021identifying}, while this approximation is very useful (and has been adopted by the LIGO-Virgo-KAGRA collaboration to make claims about precession~\cite{abac2025gw231123,abac2025gw241011}), it is only valid when the precession angle is small. Therefore, while the assumptions are valid for a large majority of observed gravitational waves, they will reach their limits precisely when the potential for a significant measurement is highest. This is also true for areas of parameter space with high mass ratios, maximal $\chi_p$ values, or when $\chi_\text{eff} \leq -0.4$ (though at least two of the three above are required).

\noindent Further, a general limitation of $\rho_p$ is its assumption of Gaussian and stationary noise. In the context of this paper, a main advantage of using evidence networks is that they do not require the two-harmonic approximation, and do not require preliminary nested sampling runs. 

\subsection{Bayes factor significance calibration}
While it is standard practice to calculate Bayes factors to make statements about precession (or other potential physics) in GW data~\cite{Hannam2022,abac2025gw231123,romero2023eccentricity,gupte2025evidence,tibrewal2026misinterpreting}, it has in practice not been possible to use them to claim the detection of precession in a real event. The main reason is that the background distribution of Bayes factors in interferometer noise is unknown. Without knowing the probability of obtaining an equally-high Bayes factor from the null hypothesis, it is impossible to evaluate the significance of a result. For example,~\cite{abac2025gw231123} finds Bayes factors upwards of $10^3$ in favour of precession in GW231123 but is unable to claim a measurement. In the case of GW200129~\cite{GWTC3} a much more modest Bayes factor of about 30 is accompanied by a much more significant claim of a measurement of precession by using the \textit{precessing SNR}~\cite{Hannam2022} (a probability of $\approx 4\cdot 10^{-5}$ for the imprints of precession in the signal to be caused by noise alone). This is also supported by~\cite{precessionRare} in the context of GWTC-3.0~\cite{GWTC3} data, where 
p-values obtained using the aforementioned \textit{precessing SNR} were compared to their expected distribution in noise for 69 events.

\begin{figure}
    \centering
    \includegraphics[width=\linewidth]{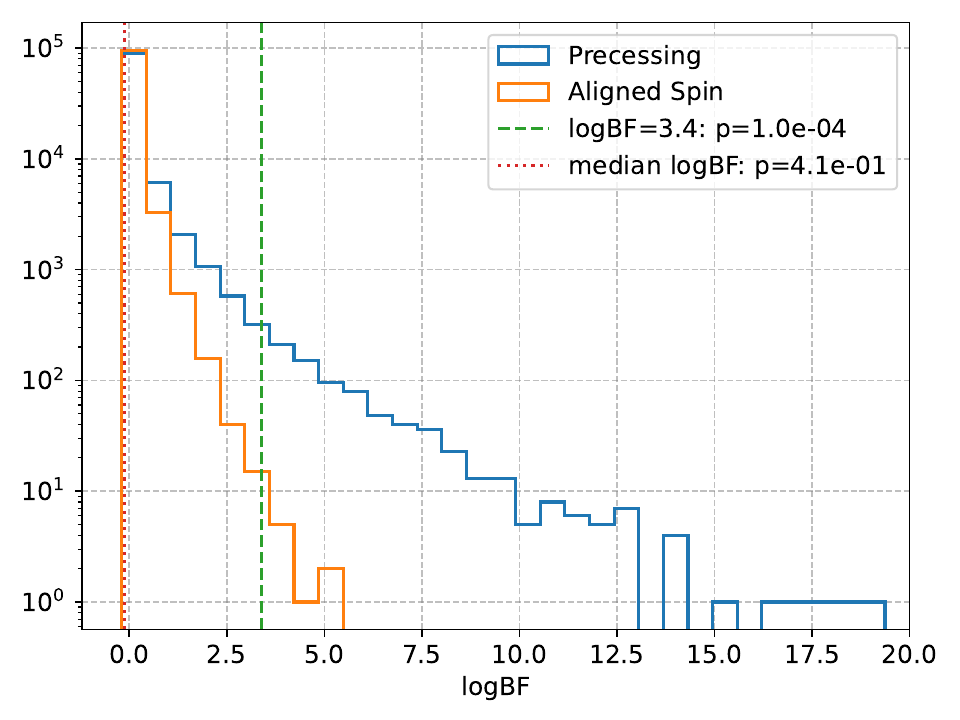}
    \caption{Distribution of Bayes factors calculated with data from scenarios I and II. As expected given the prior, most values are close to zero but the distributions can be clearly distinguished by their tails.}
    \label{fig:discussion:distr}
\end{figure}

\noindent This paper solves this problem, as the trained networks are light-weight and can be used to calculate up to $10^5$ Bayes factors per second (this includes reading data from disc and transferring the Bayes factors back to the CPU), making it trivial to construct empirical Bayes factor distributions under different model assumptions.
\Cref{fig:discussion:distr} shows such distributions for the aligned-spin and precessing models, respectively. As expected, the precessing curve has a longer tail.
While the training data used in this paper is focussed on the GW231123 prior space and can't be used to make direct statements about GW200129, in practice the priors are similar.
Thus, while any statement about the significance of precession in GW200129 is left to a future publication, when using the histogram in ~\Cref{fig:discussion:distr}, one can observe that a Bayes factor of 30 corresponds to $p = (10.0 \pm 3.9)\cdot 10^{-5}$ against the aligned-spin hypothesis.
The quoted errors were evaluated by using a Poisson likelihood to fit the ``Aligned Spin'' histogram with a power law, then generating similar p-values to obtain the Monte-Carlo uncertainty. As expected because \textit{Scenario I} assumes stationary Gaussian noise, this is compatible with aforementioned results that use the \textit{precessing SNR}~\cite{Hannam2022,precessionRare}.\\

\noindent The ability to associate an observed Bayes factor with a direct statistical significance is an important result. A more important potential result lies beyond the scope of this work: the same method can in principle be applied with no assumption of Gaussian or stationary noise.
The biggest hurdle towards this objective is tied to Bayes factors in general: their calculation assumes that the data is well described by the model. This applies here as well since extreme Bayes factors can arise when one or both models fail to accurately represent the data~\cite{pratten2020measuring}.

\subsection{Scalable sample-free population inference}
The most straightforward way to apply the methods presented above to make statements on a population scale is to modify the way that the training data is generated. Using a selection function, one can build individual populations to give as input to the networks, similarly to what was done in~\cite{jiang2025fast,leyde2026end}. The main difference in this paper is that evidence networks don't need to reconstruct multi-dimensional posterior surface distributions in order to make astrophysical statements, and are as such much smaller and faster to train. This also generally leads to more precise Bayes factor estimates (see~\cite{Evidence} and \Cref{sec:methods:val}), which should be particularly true when compared to hierarchical inference, which is known to accumulate biases for example due to Monte-Carlo approximations~\cite{talbot2023growing,heinzel2025not}.

\noindent Alternatively, as pointed out in~\cite{vitale2017fraction}, when Bayes factors test whether data comes from one of two mutually exclusive hypotheses (labelled here $a$ and $\bar{a}$), one can directly calculate the posterior on the fraction of events belonging to one model or the other. They derive:
\begin{align}\label{eq:fraction}
    \log p(f_a|\vec{d}) \propto \log p(f_a) + \sum_{k=1}^N\,&\log(f_aZ^a_k), \nonumber \\
    &+ \log\left((1 - f_a)Z^{\bar{a}}_k\right)
\end{align}
where $p(f_a|\vec{d})$ is the posterior on the fraction $f_a$ of events coming from hypothesis $a$ given the data $\vec{d}$, $p(f_a)$ is the prior on $f_a$, the index $k$ denotes a sum over all data in $\vec{d}$, and $Z^i_k$ is the evidence for model $i\in[a, \bar{a}]$ given data $d_k$. By further inserting \Cref{eq:prob}, we obtain:
\begin{align}\label{eq:fraction2}
    \log p(f_a|\vec{d}) \propto \log p(f_a) + \sum_{k=1}^N&-\log(1 + \mathcal{B}_k)\nonumber\\
    &+ \log(f_a + (1 - f_a)\mathcal{B}_k),
\end{align}
where $\mathcal{B}_k$ is the Bayes factor between models $a$ and $\bar{a}$ given data $d_k$. It is thus possible to construct population-level posterior distributions solely from event-level Bayes factors.\\

\begin{figure}[h!]
    \centering
    \includegraphics[width=0.5\textwidth]{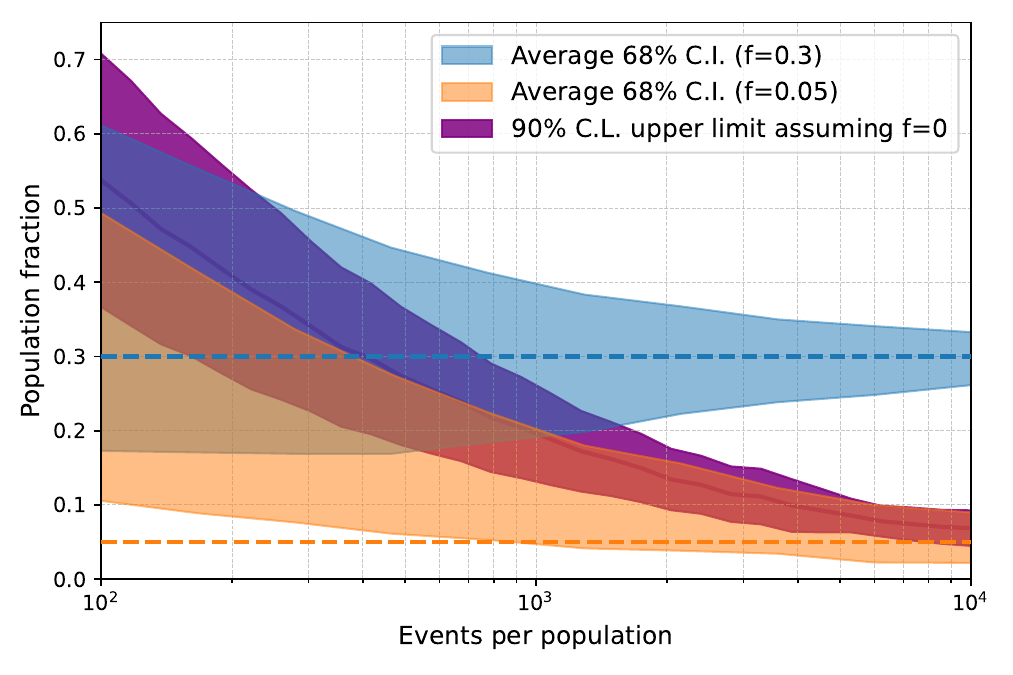}
    \caption{\textit{purple band}: $90\%$ C.L. upper limit on the precessing fraction if the injected population has $f_a = 0$ as a function of the number of events detected per population, with a $1\sigma$ uncertainty band. \textit{blue and orange}: average reconstructed fraction using \Cref{eq:fraction2} for different injected values, along with uncertainty bands. The results are biased towards the positive for low numbers of events, because the posteriors are effectively cut-off at zero when the uncertainty is higher. This is reduced for higher injected values of $f_a$ and as the number of events grows, where the distributions become Gaussian. For reference, this plot took about \SI{1}{\minute} to generate.}
    \label{fig:population:sensitivity}
\end{figure}

\noindent In \textit{Scenario I}, the models are not technically mutually exclusive because the aligned-spin parameter space is a subset of the precessing one. In practice however, the overlap between both prior spaces is small enough for \Cref{eq:fraction2} to be valid.
The amortized nature of the networks allows us to easily scale up to $10^4$ detected events per population: well beyond the numbers that the LVK is expected to measure in even the fifth observing run~\footnote{\url{https://rtd.igwn.org/projects/userguide/en/latest/capabilities.html}}. We find that the posteriors exhibit no discernable bias \textit{i.e.} the average deviations from the injected values are are of similar magnitude as the average $68\%\, C.L.$ of the posteriors. This is shown in \Cref{fig:population:sensitivity}. Since we can quickly obtain many posteriors for any injected fraction and number of events, it becomes trivial to also compute the fraction for which one could claim a detection. The results, which are compatible with~\cite{vitale2017fraction} where 200 Bayes Factors were calculated using nested sampling, are shown in \Cref{fig:population:sensitivity}, where one can see that for the current number of GW detections (about 300~\cite{GWTC4}) and assuming an injected value of zero, one could set an upper limit of about 0.35.


Recent GWTC-5 population analyses further motivate such measurements. The inferred fraction of a uniform $\chi_{\rm eff}$ component varies with primary mass: its posterior reaches approximately 0.7 at $10$--$20\,M_\odot$, decreases at intermediate masses, and rises again above $40\,M_\odot$. This component is consistent with expectations for hierarchical mergers with isotropic spin orientations~\cite{gwtc5pop,antonini2025inferring}.
These large-scale results are of course not directly comparable to the numbers presented here, since they do not consider the selection function and used data from \textit{Scenario I} in which events are drawn from a prior constructed around the posteriors measured for GW231123~\cite{abac2025gw231123}, and only for the Hanford detector.

However, this demonstrates the use of evidence networks to infer population fractions directly from strain through amortized event-level Bayes factors, without requiring intermediary posterior samples. If applied to a full catalogue, this would thus enable rapid inference and repeated full-scale injection studies. While the accuracy on realistic catalogues remains to be validated, these results establish a computationally scalable approach to population model comparison.

\begin{acknowledgements}
The author would like to thank Sama Al-Shammari, Vivien Raymond, and Stephen Green for helpful discussions and encouragement.
This work was supported by a UKRI Future Leaders Fellowship (grant number MR/Y018060/1).
The author is grateful for computational resources provided by the LIGO Laboratory and supported by National Science Foundation Grants PHY-0757058 and PHY-0823459.
This material is based upon work supported by NSF’s LIGO Laboratory which is a major facility fully funded by the National Science Foundation.
    
\end{acknowledgements}


\end{document}